\documentclass{emulateapj}
\usepackage{apjfonts}
\input{epsf}

\slugcomment{Publihed in the Astrophysical Journal Letters}

\shorttitle{Detectability of a Predicted VB~10 Mesolensing Event}
\shortauthors{L\'epine \& DiStefano}

\begin{document}

\title{On the detectability of a predicted mesolensing
  event associated with the high proper motion star
  VB~10.\altaffilmark{1}}

\author{S\'ebastien L\'epine\altaffilmark{2,3} \& Rosanne
  DiStefano\altaffilmark{4}}

\altaffiltext{1}{Based on observations made with the NASA/ESA Hubble
  Space Telescope, and obtained from the Hubble Legacy Archive, which
  is a collaboration between the Space Telescope Science Institute
  (STScI/NASA), the Space Telescope European Coordinating Facility
  (ST-ECF/ESA) and the Canadian Astronomy Data Centre (CADC/NRC/CSA).}

\altaffiltext{2}{Department of Astrophysics, American Museum of
  Natural History, Central Park West at 79th Street, New York, NY
  10024, USA; lepine@amnh.org}

\altaffiltext{3}{City University of New York, 365 Fifth Avenue New
  York, NY 10016, USA}

\altaffiltext{4}{Harvard Smithsonian Center for Astrophysics, 60
  Garden St., Cambridge MA 02138, USA; rd@cfa.harvard.edu}

\begin{abstract}
Extrapolation of the astrometric motion of the nearby low-mass star
VB~10 indicates that sometime in late December 2011 or during the
first 2-3 months of 2012, the star will make a close approach to a
background point source. Based on astrometric uncertainties, we
estimate a 1 in 2 chance that the distance of closest approach
$\rho_{min}$ will be less than 100~mas, a 1 in 5 chance that
$\rho_{min}<$~50~mas, and a 1 in 10 chance that
$\rho_{min}<$~20~mas. The latter would result in a microlensing event
with a 6\% magnification in the light from the background source, and
an astrometric shift of 3.3~mas. The lensing signal will however be
significantly diluted by the light from VB~10, which is 1.5~mag
brighter than the background source in B band, 5 mag brighter in I
band, and 10~mag brighter in K, making the event undetectable in all
but the bluer optical bands. However, we show that if VB~10 happens to
harbor a $\sim1$ M$_{J}$ planet on a moderately wide
($\approx0.18$~AU$-0.84$~AU) orbit, there is a chance (1\% to more
than 10\%, depending on the distance of closest approach and orbital
period and inclination)  that a passage of the planet closer to the
background source will result in a secondary event of higher
magnification. The detection of secondary events could be made
possible with a several-times-per-night multi-site monitoring
campaign.
\end{abstract}

\keywords{gravitational lensing: micro \--- astrometry \--- stars:
  low-mass \--- stars: planetary systems}

\section{Introduction}

Microlensing monitoring programs have been operating since the early
1990s
\citep{Paczynski.1995,Udalski.etal.1997,Palanque-Delabrouille.etal.1998,Alcock.etal.2000,Bond.etal.2001}. These
programs monitor large areas of the sky in the hope that a planetary
or stellar mass object will pass close enough to the position of a
more distant ``source'' star to cause a detectable lensing event. More
than 8,000 lensing event candidates have now been identified. Among all
possible gravitational lenses, nearby masses (d$\lesssim$~1 kpc) hold a
special place \citep{DiStefano.2008a,DiStefano.2008b}. Their 
Einstein radii ($\theta_e$) are larger than they would be were they to
be as distant as the typical microlens (several kpc away along
directions to the Bulge, and several tens of kpc away along directions
to the Magellanic Clouds), and their angular motions are significantly
faster. The combination of a large Einstein ring and angular speed
means that nearby stars have a larger probability of producing a
detectable lensing event. Furthermore, because nearby stars can be
independently detected, the degeneracy inherent in
lensing light curves can be overcome, allowing us to determine their
gravitational mass. Because many nearby stars already have their
positions and proper motions known to some accuracy, we can {\sl
  predict} and plan for the monitoring of specific events
\citep{Paczynski.1995,Gould.2000,Salim.Gould.2000,DiStefano.2008a,DiStefano.2008b}. Very
nearby stars moving against the backdrop of the Milky Way
\citep{Lepine.etal.2002,Lepine.2005,Lepine.2008,Rattenbury.Mao.2008}
make promising targets in this regard because the high field densities
provide more opportunities for chance alignments.

\begin{figure*}[t]
\epsscale{0.85}
\plottwo{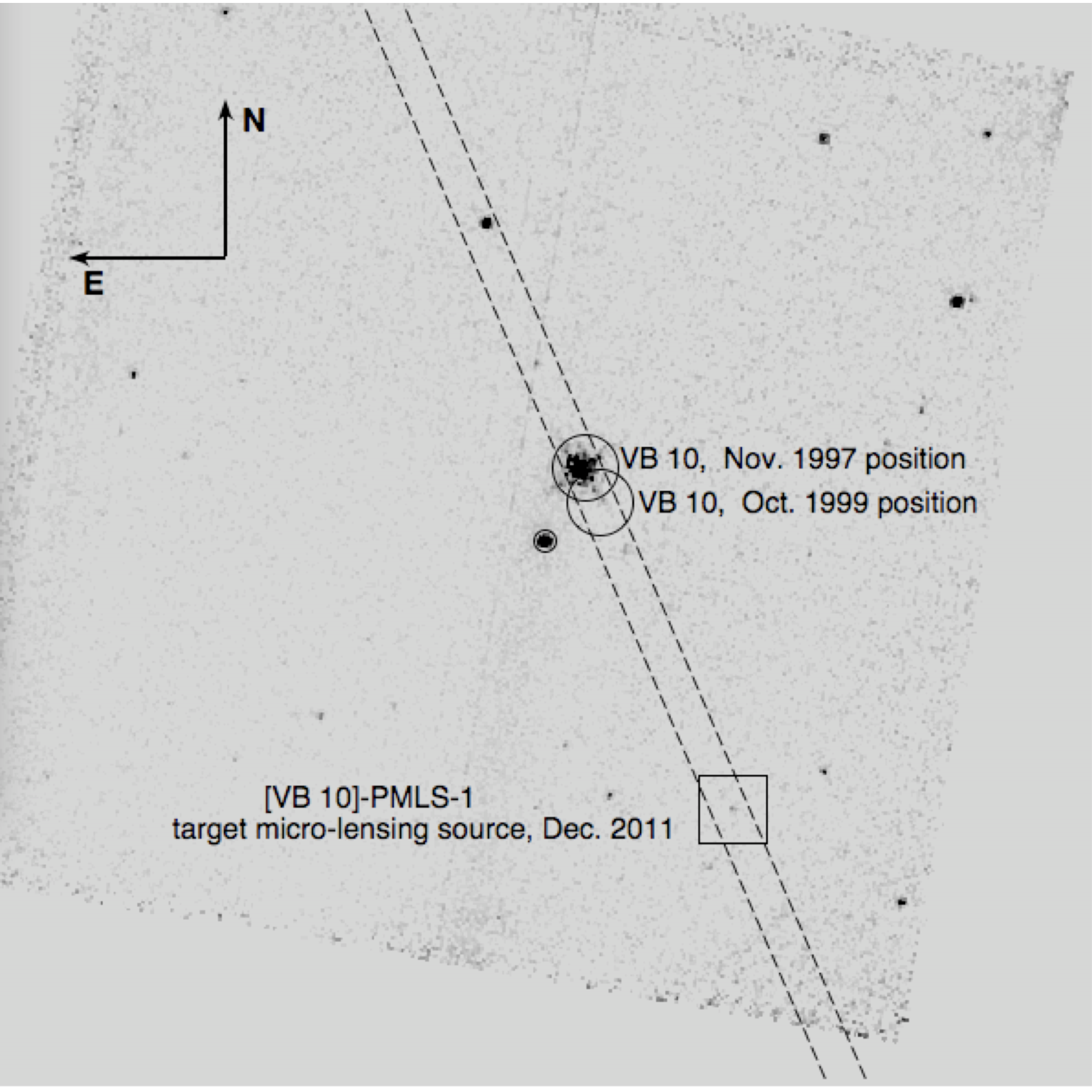}{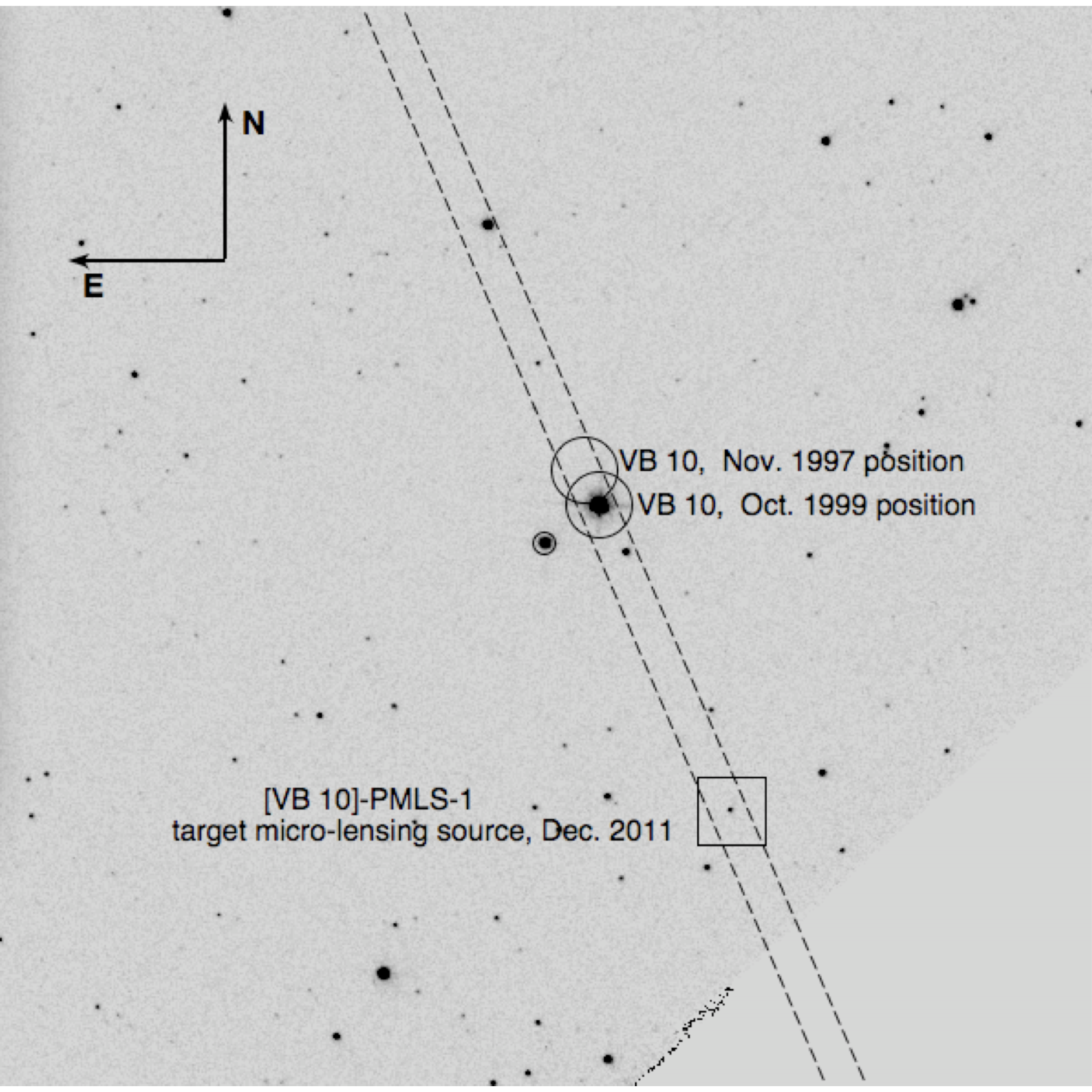}
\caption{Hubble Space Telescope NICMOS (left) and WFPC2 (right) images
  of the field around the nearby low-mass star VB~10, from the Hubble
  Legacy Archive. The frame is 1\arcmin$\times$1\arcmin, with North up
  and East left. VB~10 is labeled and identified with an open
  circle. Two dashed lines running parallel to the measured proper
  motion of VB~10 are overlaid. The predicted path is taking the star
  to within a short angular distance of a background source,
  identified as [VB~10]-PMLS-1. We predict a lensing event to occur in
  late 2011 or early 2012.}
\end{figure*}

Lensing events can also be used to search for stellar or
substellar companions. Because the shape of the lensing
light-curve is sensitive to the mass distribution of the lens
system
\citep{Mao.Paczynski.1991,Gould.Loeb.1992,DiStefano.Scalzo.1999a,DiStefano.Scalzo.1999b},
orbiting planets can produce significant, short-lasting deviations
from the standard microlensing profile. To detect a planet in a
lensing light curve, moderate-cadence monitoring must be supplemented
by high cadence observations
\citep{Gould.Loeb.1992,Griest.Safizadeh.1998,DiStefano.Scalzo.1999a,DiStefano.Scalzo.1999b},
triggered when a deviation to the single-mass microlensing light curve
is detected. Several exoplanets have already been detected using this
method, which holds strong potential to provide a wealth of
statistical data on planetary masses and orbital distributions,
especially in the wide separation range. Unfortunately, most
planetary systems detected through microlensing are relatively far
from the Sun, and their host stars are much dimmer than the background
source \citep{Janczak.etal.2010,Sumi.etal.2010,Miyake.etal.2011}. This
seriously limits the possibilities for follow-up observations,
making it near-impossible to confirm the presence of the
planet by other means (spectroscopic, photometric, astrometric) and
limiting the information one can obtain about the host star. Most
exoplanet detections through microlensing light-curves are essentially
``one shot deals''.

Microlensing by a nearby star, also called ``mesolensing'', would have
all the potential for detecting exoplanets through microlensing, with
the added bonus that the lens system would be much more amenable to
follow-up observations. The presence of the planet could potentially
be confirmed using secondary detection methods, and the properties of
the host star could be easily determined, if not already known. 

In this Letter, we predict a future event associated with the
low-mass star VB~10, based on precise astrometric measurements made with the
{\sl Hubble Space Telescope (HST)}. We investigate the detectability
of the event itself and of any secondary event that might be produced
by a hypothetical planet in orbit around the star.

\begin{deluxetable}{lrr}
\tabletypesize{\scriptsize}
\tablecaption{Ancillary data}
\tablewidth{240pt}
\tablehead{
\colhead{Datum} &
\colhead{Value} &
\colhead{Source} 
}
\startdata
\cutinhead{VB 10 (lens)}
\hline
R.A.(ICRS)   &  19 16 57.622 & \citet{cutri.2003}\\
Decl.(ICRS)  & +05 09 02.18  & \citet{cutri.2003}\\
Epoch\tablenotemark{a} &  1999.58 &  \citet{cutri.2003}\\
$\mu_{R.A.}$  &   589.01$\pm$0.2 mas yr$^{-1} $& \citet{Pravdo.Shacklan.2009}\\
$\mu_{Decl.}$ & -1361.12$\pm$0.3 mas yr$^{-1}$ & \citet{Pravdo.Shacklan.2009}\\
$\pi_{trig}$  &   171.6$\pm$1.4 mas           & \citet{Pravdo.Shacklan.2009}\\
B & 19.5 mag& this paper \\
V & 17.2 mag& \citet{Casagrande.etal.2008}\\
I & 12.8 mag& \citet{Casagrande.etal.2008}\\
J & 9.90 mag& \citet{cutri.2003}\\
H & 9.23 mag& \citet{cutri.2003}\\
K & 8.76 mag& \citet{cutri.2003}\\
\cutinhead{[VB~10]-PMLS-1 (lensed source)}
\hline
R.A.(ICRS)   &  19 16 57.136 & this paper\\
Decl.(ICRS)  & +05 08 45.34  & this paper\\
B & 21.0 mag& this paper \\
I & 18.2 mag& this paper \\
H & 16 mag& this paper \\
\enddata
\tablenotetext{a}{Epoch of the R.A. and Decl. measurements.}
\end{deluxetable}

\section{Hubble Space Telescope archival data on VB~10}

The low mass star VB~10 (= V 1298 Aql), discovered by
\citet{VanBiesbroek.1961}, is a common proper motion companion to the
high proper motion star Ross 652 (= V 1428 Aql). The pair is separated
by $\approx74^{\prime\prime}$, share a proper motion $\mu\simeq$1.5
$\arcsec$ yr$^{-1}$, and is located at a distance $d=5.8$pc. The
distance and angular separation correspond to a projected physical
separation $\approx429$ AU in the plane of the sky. The two
stars are currently moving against a rich Milky Way background at a
Galactic latitude $b=+7.86$. The proper motion and parallax of VB~10
has recently been measured to a high level of accuracy by
\citet{Pravdo.Shacklan.2009}. Their claimed detection of a 6 M$_{J}$
exoplanet orbiting VB~10 has however not been confirmed by subsequent
studies \citep{Bean.etal.2010,Lazorenko.etal.2011}. Radial velocity
measurements of VB~10 to a precision of 150 m s$^{-1}$ have shown no
evidence for a companion \citet{AngladaEscude.etal.2010}. Combined
with the astrometry from \citet{Pravdo.Shacklan.2009}, these radial
velocity measurements set an upper limit of 2.5 M$_{J}$ for a planet
orbiting VB~10 with an orbital period $<1$ year.

VB~10 was imaged by {\it HST} on two occasions with the {\it Near
 Infrared Camera and Multi-Object Spectrograph} (NICMOS). Exposures
in the F113N filter were obtained on 1997 November 19, and in the
F187N filter on 1998 June 19. The images are available from the Hubble
Legacy Archive (HLA), where they have been combined into a single
cleaned image with the {\rm drizzle} routine, which also performs
astrometric corrections for field distortions. A section of the
resulting F187N image is shown in Figure 1.

VB~10 was also imaged with the {\it Wide Field and Planetary Camera 2}
(WFPC2) on 1999 October 2. Exposures were obtained in the F439W,
F555W, F775W and F814W filters. The images are also available from the
HLA, where they too have been combined with {\rm drizzle}. A section
of the composite F814W image is shown in Figure 1. All the NICMOS and
WFPC2 images in the HLA have been remapped on a grid with angular
scale 0.1\arcsec pixel$^{-1}$, and with the X-axis oriented along
Right Ascension and the Y-axis along Declination.

The proper motion of VB~10 is apparent from a comparison of the NICMOS
and WFPC2 frames, obtained two years apart. The location of VB~10 is
noted in Figure 1 (circles) and a pair of dashed lines running
parallel to its proper motion are overlaid. A close examination of the
F187N and F814W images reveals a faint source lying along the
projected path of VB 10. That source is identified by a box. The
source is detected in the F439W, F555W and F775W frames as well,
but fell outside the field-of-view of the NICMOS F113W exposures. We
name the source [VB~10]-PMLS-1, for VB~10 Predicted Mesolensing Source
number 1.

Photon counts show that the source is 1.5 mag fainter than VB~10 in
F439W (equivalent to $B$ band), 5.4 mag fainter in F814W ($I$ band),
and $\approx$~7~mag fainter in F187N (close to infrared $H$
band). VB~10 has visual magnitudes V=17.2 and I=12.8
\citep{Casagrande.etal.2008}, and the 2MASS catalog lists the star
with infrared magnitude H=9.23. The blue magnitude is estimated at
B=19.5 from the WFPC2 images. This puts [VB~10]-PMLS-1, the background
source, at B$\simeq$21.0, I$\simeq$18.2 and H$\approx$16. Data on VB
10 and the background source are summarized in Table 1.

Limits can be placed on the relative proper motion of [VB~10]-PMLS-1
by comparing the position of the source on the NICMOS and WFPC2
images. We find that the position in the WFPC2 and NICMOS epochs agree
to within 26 mas, which indicates a proper motion $<14$ mas
yr$^{-1}$. Random field stars tend to have proper motions smaller than
this, however, with typical values in the 5-10 mas yr$^{-1}$ range. A
search of the SDSS catalog \citep{Adelman.2008} for example, turns up
five times as many $r<20$ stars with proper motions $\mu<10$~mas
yr$^{-1}$ than stars with $\mu>10$~mas yr$^{-1}$. If [VB~10]-PMLS-1
has any net proper motion, it is more likely to be in the 5-10~mas
yr$^{-1}$ range.



\section{Predicted primary mesolensing event}

\begin{figure}[t]
\epsscale{2.05}
\plotone{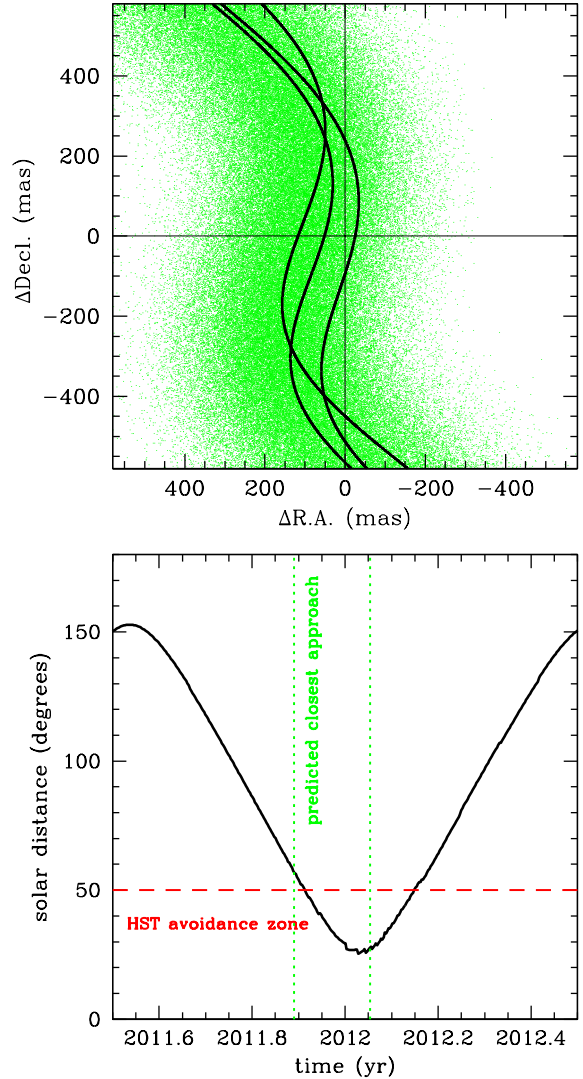}
\caption{Top: predicted range of trajectories of the star VB~10 in the
  reference frame of the target mesolensing source
  [VB~10]-PMLS-1. The point density is proportional to the probability
  of a path passing by that location, given the uncertainties in the
  astrometric solution for the motion VB~10 and proper
  motion of the background source. Three possible trajectories are shown, with
  distances of closest approach $\approx$100~mas, 50~mas, and
  20~mas; these are used for the simulations shown in Figures 3 and 4.
  Bottom: angular distance to the Sun of VB~10 around the predicted
  time of the mesolensing event. The star is likely to be too close to
  the Sun to be observed with HST, and will be challenging even for
  ground-based observations. The vertical dotted lines show the most
  likely range for the peak of the primary lensing event. (Secondary,
  planet-lens events may occur earlier or later.)}
\end{figure}

To calculate the time and distance of closest approach to the
background source, we extrapolate the parallax and
proper motion of VB~10 based on the astrometric solution of
\citet{Pravdo.Shacklan.2009}. We use the drizzled WFPC2 images to
calculate the position of VB~10 in the reference frame of
[VB~10]-PMLS-1 at the epoch of the 1999 {\it HST} visit.

To account for astrometric errors and uncertainties in the proper
motion of [VB~10]-PMLS-1, we perform Monte-Carlo simulations to
generate 10$^4$ possible paths for VB~10, with variations in its
parallax and proper motion within the range of astrometric
uncertainties, and with the assignment of random components of proper
motion [VB~10]-PMLS-1 with mean values
($\bar{\mu_{RA}}$,$\bar{\mu_{DEC}}$)=(0,0)~mas yr$^{-1}$ and
dispersions ($\sigma_{\mu_{RA}}$,$\sigma_{\mu_{RA}}$)=(10,10)~mas
yr$^{-1}$. Results are illustrated in Figure 2, where the point
density is proportional to the probability of a path passing by that
location. Statistics show that encounters with distance at closest
approach $\rho_{min}<$100~mas occur in 44\% of trajectories,
encounters with $\rho_{min}<$50~mas in 22\% of trajectories,
while encounters with $\rho_{min}<$20~mas occur 9\% of the time. Three
trajectories with $\rho_{min}\approx$(20,50,100)~mas are shown in
Figure 2 as examples.

To investigate the effects of a smaller proper motion for
[VB~10]-PMLS-1, we generate an additional set of paths using
($\sigma_{\mu_{RA}}$,$\sigma_{\mu_{RA}}$)=(5,5)~mas yr$^{-1}$. We then
find trajectories with $\rho_{min}<$100~mas occuring 51\% of the
time, trajectories with $\rho_{min}<$50~mas 22\% of the time, and
trajectories with $\rho_{min}<$20~mas 8\% of the time. Given the
similar results, we conclude that under the assumption that
[VB~10]-PMLS-1  has a proper motion in the 5-10~mas yr${-1}$ range,
there is a roughly 1 in 2 chance that an alignment will occur with
separation of 100~mas or less, a 1 in 5 chance for an alignment with
separation of 50~mas, and a 1 in 10 chance of an alignment with
separation 20~mas or less. The time of closest approach varies
depending on the relative path of VB~10 to the background sources. The
median value is 2011 December 21, with a dispersion of $\pm$30 days.

From the projected trajectory of VB~10, and assuming the M8.0V dwarf
to have a mass of $0.08$M$_{\odot}$ \citep{Baraffe.Chabrier.1996}, it is
possible to predict the magnification and astrometric shift of
[VB~10]-PMLS-1 due to gravitational lensing. The angular radius of the
Einstein ring ($\Theta_e$) of VB~10 is approximately equal to
$10$~mas. As discussed
above, trajectories consistent with the astrometric measurements
produce distances of closest approach from $\rho_{min}\approx2\Theta_e$
($20$~mas) to $\rho_{min}\gtrsim10\Theta_e$ ($>100$~mas). At $20$~mas,
the total magnification of the background star is $6\%$. The
magnification at $50$~mas is $0.3\%$, while it is $\sim 0.02\%$ at
$100$~mas. The decline is steep, because the magnification falls as
$\rho^{-4}$. The centroid shift falls off more slowly with distance
($\sim \rho^{-1}$), and the astrometric shift would be approximately
($3.3$,$1.9$,$1.0$)~mas, for distances of closest approaches of
($20$,$50$,$100$)~mas, respectively.

There are complications to observing this predicted mesolensing
event. First, VB~10 is brighter than [VB~10]-PMLS-1. The magnification
light curve will thus be significantly diluted, and the astrometric
shift reduced. In the infrared ($J$) the $>600$ dilution factor will
reduce the amplitude of a 6\% event to less than 1 part in 10$^4$. In
the blue ($B$) however, the light from [VB~10]-PMLS-1 is diluted
by a factor $\sim5$, which would reduce the amplitude of the event to
a $\approx1$\% of the integrated flux. In addition, VB~10 will be
close to the Sun in December and January, making ground observations
challenging and space observations impossible when the source falls
e.g. within the Solar avoidance zone of HST (Figure 2, bottom panel).

\begin{figure}[t]
\epsscale{2.05}
\plotone{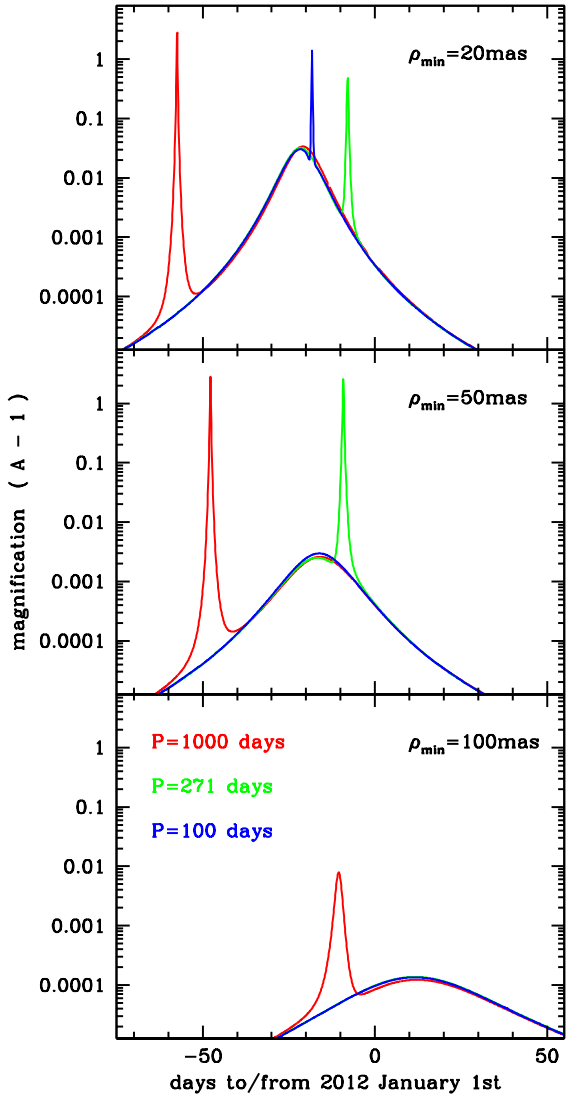}
\caption{Magnification curves for three of the best case
  scenarios from our simulations, where a planet orbiting VB~10 causes
  a significant secondary microlensing event. The same three orbiting
  planets are shown for different possible trajectories of VB~10 relative
  to the source, yielding different distances of closest approaches (20
  mas, 50 mas, and 100 mas, top to bottom.) Each colored light curve
  corresponds to a system hosting a
  hypothetical 1 M$_{J}$ planet with an orbit of 1000 days (red), 271
  days (green), and 100 days (blue). Planets on shorter orbits only
  produce notable magnifications in short-distance encounters, while
  planets with larger orbital separations can produce significant
  magnifications events even when the distance of closest approach
  between VB~10 and the background source is large.}
\end{figure}

\section{Secondary lensing event from a planetary companion}

If VB~10 has a dark companion, the effects of lensing can be
enhanced. \citet{Lazorenko.etal.2011} measurements exclude a 3.2
M$_{J}$ planet with period of 270 days, and the
\citet{AngladaEscude.etal.2010} radial velocity measurements exclude
most planets with masses $>$2.5 M$_{J}$. However planets with a range
of masses and orbits are still possible.

The type of secondary effect expected in the microlensing light curve
depends on the orbital separation between the planet and VB~10, but
also on the orientation of the orbital plane, and on the orbital
phase. We generate sets of microlensing light curves for one of
three possible trajectories for VB~10, with distances of closest
approach $\rho_{min}$=~20~mas, 50~mas, and 100~mas; these are the
trajectories shown in Figure 2. For each path, we calculate circular
face-on planetary orbits in which a Jupiter-mass planet orbits
VB~10. We consider three separate orbital
periods of 100~days, 271~days, and 1,000~days. We compute light curves
for each of several thousand choices of the initial orbital phase,
chosen uniformly between $0$ and $2\, \pi$. This allows us to compute
the fraction of the time the peak magnification is larger than some
arbitrary value (e.g. 30\%).

We find a small fraction of events in which the planet makes a
significantly closer approach to the source star than does VB~10
itself, resulting in a large magnification (Figure 3). For the
$\rho_{min}= 20$~mas path, events of all three orbital periods are
found to generate some high magnification peaks. For $\rho_{min}=
50$~mas trajectories, only the planets on $271$-day, and $1,000$-day
orbits generate high magnification events, while for $\rho_{min}=
100$~mas trajectories, only planets on a $1,000$-day orbit generate
events with significant magnification. For planets on a $1,000$-day
orbit, we found the probability of a $> 30\%$ magnification event to
be $\approx 0.7\%$ for all three values of $\rho_{min}$. Planets on a
$271$-day orbit had a $\approx 2\%$ probability of a generating $>
30\%$ magnification events in both $\rho_{min} = 20$~mas and
$\rho_{min} = 50$~mas models (no such events are generated in the
$\rho_{min} = 100$~mas case). The direction of orbital motion
influences the event probability for planets with shorter orbital
periods. For example, a planet on a $100$-day orbit (which can
generate events for $\rho_{min}$ less than or equal to about $20$~mas)
has a $12\%$ chance of generating events with $> 30\%$ magnifications
with one orbital direction and a $5\%$ chance if the orbital direction
is reversed, giving an average of $8.5\%$.

In addition, we find that secondary events can occur up to 3
months before or after the primary event, depending on the period and
orbital phase of the companion. Predicted composite light curves for
the magnification events shown in Figure 3 are displayed in Figure
4, and show the apparent magnitude of the unresolved source and lens
accounting for the dilution from the light of VB~10. The secondary
events retain symmetrical profiles, but blending reduces the timescale
of the peak to $<$1~day.

\section{Discussion and conclusions}

\begin{figure}
\epsscale{2.4}
\plotone{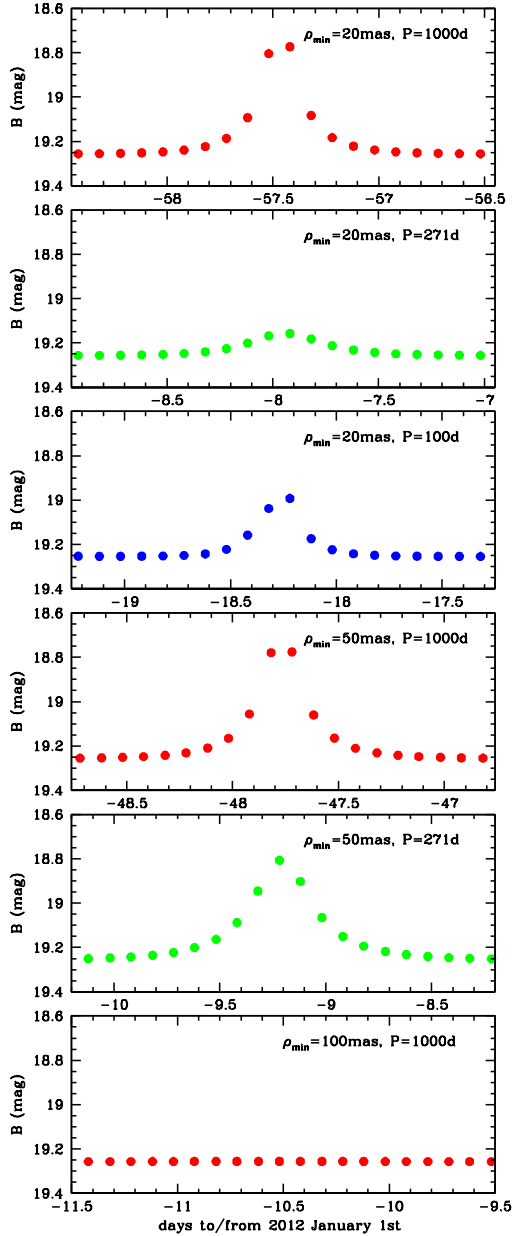}
\caption{Composite light curves of VB~10 and the unresolved background
  source, for the best-case scenarios depicted in Figure 3. Similar
  events, though they would have only a small chance of occurring,
  would be easily detectable with modest photometric precision, even
  as the primary event would remain undetectable. Any intrinsic
  variability in VB~10, due to e.g. flares, would be superposed on
  these lensing light-curves. Flares would typically occur on
  timescales of ~1 hour and have asymmetric profiles.}
\end{figure}

This prediction of a microlensing event associated with the very
low mass star VB~10 constitutes a unique opportunity for observational
astronomy. Detection of the primary event presents severe challenges.
The lensed source is a faint background star, which means that the
event will suffer significant dilution from VB~10. The event
will also occur at an inopportune time, because VB~10 will be close to
the Sun near the time of the event, making it an evening star
at unfavorable airmass. Proximity to the Sun will also prevent
observations from {\it HST} and other space telescopes, which have
solar avoidance angles of $\sim$50 degrees. In the red or infrared,
the large dilution from the VB 10 light will make the event
undetectable using existing facilities. At best, the light-curve for the
unresolved source and lens may reach an amplitude $\sim1$\% in
the blue, for a source with magnitude $B=19.5$. The detection of such
a weak event would be difficult if not impractical.

However, our simulations show that intensive monitoring of the event
in the months {\em before} and {\em after} the primary event could
potentially yield the microlensing detection of a planet orbiting
VB~10, if the proper alignment conditions are met. Evidence has now
been accumulating that low-mass stars can hold planetary
retinues. Indeed the 0.3$M_{Sun}$ M dwarf GJ 581 is now the system
with the largest number of confirmed planets \citep{Mayor.etal.2010},
and several other low-mass M dwarfs have been found to have planetary
companions. A massive planet was also recently discovered orbiting an
M dwarf through microlensing \citet{Batista.etal.2011}. It is thus a
definite possibility that VB~10 might have associated planets. If the
proper alignment conditions are met, then secondary microlensing events
could reach high enough magnifications to be detected. If an event
yields a total magnification of the source+lens light by 30\% or more,
then a signal to noise ratio S/N$>$10 would be sufficient for a
3-$\sigma$ detection achievable in minutes of exposure on a 1-meter
class telescope. 

Simulations show that high magnification events due to orbiting
planets would have timescales of several hours. A monitoring campaign
on VB~10 to detect them would thus require intensive monitoring
from multiple sites of various longitudes. These events could occur at
any time up to a few months before and after the peak of the primary
event, which means that they could be occuring at anytime through
winter 2012 or even later.

Intrinsic variability in VB~10 might however complicate the detection
of a planetary microlensing event. VB~10 is a known flare star
\citep{Berger.etal.2008}, and the activity and flaring rate for a red
dwarf with very late spectral type (M8) is expected to be high
\citep{West.etal.2008,Hilton.etal.2010}. Flares, however, typically
show a rapid rise on timescales of minutes, followed by an exponential
decay lasting $<$1 hour \citep{Welsh.etal.2007}. This is in contrast
to the symmetric light curve with a timescale of severeal hours
expected of a hypothetical planetary microlensing event (see Figure
4). It will be critical to characterize the base variability level and
intrinsic variability patterns in VB~10, which will be possible when
the system is no longer susceptible to microlensing.

\acknowledgments

{\bf Acknowledgments}

We wish to thank James Matthews for help with the calculations and
figures. We also thank the anonymous referee for providing useful and
constructive suggestions.

This research was supported at AMNH by grants AST 06-07757 and AST
09-08419 from the United States National Science Foundation
(NSF). Support was also provided in part by NSF grants AST 07-08924
and AST 09-08878 at the CfA.



\begin{thebibliography}{}


\bibitem[Adelman-McCarthy et al.(2009)]{Adelman.2008}
Adelman-McCarthy, J.K. et al. 2009, \apjs, 182, 543

\bibitem[Alcock et al.(2000)]{Alcock.etal.2000}
Alcock, et al. 2000, \apj, 542, 281

\bibitem[Anglada-Escud\'e et al.(2010)]{AngladaEscude.etal.2010}
Anglada-Escud\'e, et al. 2010, \apj, 711, L24

\bibitem[Batista et al.(2011)]{Batista.etal.2011}
Batista, V., et al. 2011, \aap, 529, 102

\bibitem[Baraffe \& Chabrier(1996)]{Baraffe.Chabrier.1996}
Baraffe, I., \& Chabrier, G. 1996, \apj, 461, L51

\bibitem[Bean et al.(2010)]{Bean.etal.2010}
Bean, Pravdo, E., \& Shacklan 2009, \apj, 700, 623 

\bibitem[Berger et al.(2008)]{Berger.etal.2008}
Berger, E., et al. 2008, \apj, 676, 1307

%

\bibitem[Bond et al.(2001)]{Bond.etal.2001}
Bond, I. A., et al. 2001, \mnras, 327, 868

\bibitem[Casagrande et al.(2008)]{Casagrande.etal.2008}
Casagrande, L., Flynn, C., \& Bessell, M. 2008, \mnras, 389, 585

\bibitem[Cutri, et al.(2003)]{cutri.2003}
Cutri, R. M., et al. 2003, 2MASS Point Source Catalogue, University of
Massachusetts and Infrared Processing and Analysis Center,
(IPAC/California Institute of Technology)

\bibitem[DiStefano(2008a)]{DiStefano.2008a}
Di Stefano, R., 2008a, \apj, 684, 46

\bibitem[DiStefano(2008b)]{DiStefano.2008b}
Di Stefano, R., 2008b, \apj, 684, 59

\bibitem[DiStefano \& Scalzo(1999a)]{DiStefano.Scalzo.1999a}
Di Stefano, R., \& Scalzo, R. A. 1999a, \apj, 512, 564

\bibitem[DiStefano \& Scalzo(1999b)]{DiStefano.Scalzo.1999b}
Di Stefano, R., \& Scalzo, R. A. 1999b, \apj, 512, 579



%


\bibitem[Gould(2000)]{Gould.2000}
Gould, A. 2000,  \apj, 532, 936

\bibitem[Gould \& Loeb(1992)]{Gould.Loeb.1992}
Gould, A., \& Loeb, A. 1992, \apj, 396, 104

\bibitem[Griest \& Safizadeh(1998)]{Griest.Safizadeh.1998}
Griest, K., \& Safizadeh, N. 1998, \apj, 500, 37

\bibitem[Hilton et al.(2010)]{Hilton.etal.2010}
Hilton, E. J., West, A. A., Hawley, S. L., \& Kowalski, A. F., 2010,
AJ, 140, 1402


\bibitem[e.g. Janczak et al.(2010)]{Janczak.etal.2010}
Janczak, J., et al. 2010, \apj, 711, 731



\bibitem[Lazorenko et al.(2011)]{Lazorenko.etal.2011}
Lazorenko, P. F., et al. 2011, \aap, 527, 25

\bibitem[L\'epine, Shara, \& Rich(2002)]{Lepine.etal.2002}
L\'epine, S. Shara, M. M., \& Rich, R. M. 2002, \aj, 124, 1190

\bibitem[L\'epine(2005)]{Lepine.2005}
L\'epine, S. 2005, \aj, 130, 1247

\bibitem[L\'epine(2008)]{Lepine.2008}
L\'epine, S. 2008, \aj, 135, 2177

\bibitem[Mao \& Paczynski(1991)]{Mao.Paczynski.1991}
Mao, S., \& Paczynski, B. 1991, \apj, 374, 37


\bibitem[Mayor et al.(2010)]{Mayor.etal.2010}
Mayor, M., et al. 2010, \aap, 507, 487

\bibitem[Miyake et al.(2011)]{Miyake.etal.2011}
Miyake, N., et al. 2010, \apj, 728, 120

\bibitem[Paczynski(1995)]{Paczynski.1995}
Paczynski, B. 1995, Acta Astronomica, 45, 345

\bibitem[Palanque-Delabrouille et
  al.(1998)]{Palanque-Delabrouille.etal.1998}
Palanque-Delabrouille, N., et al. 1998, \aap, 332, 1

\bibitem[Pravdo \& Shacklan(2009)]{Pravdo.Shacklan.2009}
Pravdo, E., \& Shacklan 2009, \apj, 700, 623 

\bibitem[Rattenbury \& Mao(2008)]{Rattenbury.Mao.2008}
Rattenbury, N. J., \& Mao S. 2008, \mnras, 385, 905

\bibitem[Salim \& Gould(2000)]{Salim.Gould.2000}
Salim, S., \& Gould, A. 2000, \apj, 539, 241

\bibitem[Sumi et al.(2010)]{Sumi.etal.2010}
Sumi, T., et al. 2010, \apj, 710, 1641

\bibitem[Udalski et al.(1997)]{Udalski.etal.1997}
Udalski, A., Kubiak, M., \& Szymanski, M. 1997, AcA, 47, 319

\bibitem[Van Biesbroeck(1961)]{VanBiesbroek.1961}
Van Biesbroeck, G. 1961, \aj, 66, 528

\bibitem[Welsh et al.(2007)]{Welsh.etal.2007}
Welsh, B. A., et al. 2007, \apjs, 173, 673

\bibitem[West et al.(2008)]{West.etal.2008}
West, A. A., et al. 2008, \apj, 135, 785

\end{thebibliography}
\end{document}